# Developing Machine-Learned Potentials for Coarse-Grained Molecular Simulations: Challenges and Pitfalls

Eleonora Ricci*

*Institute of Informatics and Telecommunications & Institute of Nanoscience and Nanotechnology, National Centre for Scientific Research "Demokritos", Athens, Greece,* e.ricci@inn.demokritos.gr

George Giannakopoulos

*Institute of Informatics and Telecommunications, National Centre for Scientific Research "Demokritos", Athens, Greece & SciFY P.N.P.C., Greece,* ggianna@iit.demokritos.gr

Vangelis Karkaletsis

*Institute of Informatics and Telecommunications, National Centre for Scientific Research "Demokritos", Athens, Greece,* vangelis@iit.demokritos.gr

Doros N. Theodorou

*School of Chemical Engineering, National Technical University of Athens, Athens, Greece,* doros@chemeng.ntua.gr

Niki Vergadou*

*Institute of Nanoscience and Nanotechnology, National Centre for Scientific Research "Demokritos", Athens, Greece,* n.vergadou@inn.demokritos.gr

**ABSTRACT**

Coarse graining (CG) enables the investigation of molecular properties for larger systems and at longer timescales than the ones attainable at the atomistic resolution. Machine learning techniques have been recently proposed to learn CG particle interactions, i.e. develop CG force fields. Graph representations of molecules and supervised training of a graph convolutional neural network architecture are used to learn the potential of mean force through a force matching scheme. In this work, the force acting on each CG particle is correlated to a learned representation of its local environment that goes under the name of SchNet, constructed via continuous filter convolutions. We explore the application of SchNet models to obtain a CG potential for liquid benzene, investigating the effect of model architecture and hyperparameters on the thermodynamic, dynamical, and structural properties of the simulated CG systems, reporting and discussing challenges encountered and future directions envisioned.

**CCS CONCEPTS** • Applied computing → Physical sciences and engineering • Computing methodologies → Machine learning → Machine learning approaches → Neural Networks • Computing methodologies → Modeling and simulation → Simulation types and techniques → Molecular simulation

**Additional Keywords and Phrases:** Coarse-graining, Molecular Simulations, Machine Learning, Neural Network Potential, Hierarchical Modelling

---

* To whom correspondence should be addressed.

# 1 INTRODUCTION

Machine learning (ML) is having increasing impact in the physical sciences, engineering, and technology, addressing research problems that range from molecular reaction mechanisms to high-throughput screening of functional materials.

One strategy for representing molecules mathematically is through the use of graphs, whose nodes and edges correspond to atoms and bonds or interatomic distances, respectively. By performing multiple convolution operations on a graph, each node can influence other, increasingly distant, nodes. The use of graph neural networks has recently shown great promise in the development of improved atomistic force fields, trained on quantum mechanical calculations [1]. On the other hand, the implementation of ML for the generation of coarse grained (CG) mapping schemes [2], [3], and CG force fields required for developing hierarchical multiscale modelling schemes [4] on the basis of atomistic simulations is a less explored topic, and their application to the study of complex bulk systems is still rare [2],[5].

A main advantage of ML models in the context of molecular simulations is that they are endowed with higher flexibility and expressive character compared to traditional CG models. Thus, they are potentially more capable of representing complex energy hypersurfaces and capturing many-body interactions that cannot be accurately modelled with traditional CG force fields using predefined mathematical functions.

In this work, we adopted Graph Convolutional Neural Network (GCNN) architectures, as proposed by Schütt et al. [5], [6], to develop CG Machine Learned potentials for bulk liquid systems, implementing a strategy that includes a force-matching scheme [7]. We analyzed the effect of model size and hyperparameters on the thermodynamic, structural, and dynamical properties of benzene liquid system. The effect of these aspects on the simulations performed with a resulting NN CG potential is rarely discussed in the existing literature. However, in several cases, unphysical behavior is observed, both in terms of structure and of dynamics of the systems, that could not be straightforwardly correlated with the accuracy achieved during the training process. The effect of model size, hyperparameters, and loss function definitions was explored. The obtained results and observations can serve as a constructive reference of future works for the identification of challenges and difficulties in the domain of ML-based CG force field development and inspire additional mitigation measures.

# 2 METHODOLOGY

## 2.1 Force Matching

In this work, a force matching scheme has been adopted for the determination of the CG potential. This method, introduced by Ercolessi and Adam [8], involves the minimization of the difference between the atomistic forces, $\mathbf{F}^A \in \mathbb{R}^{3n}$ (where n is the number of atoms), projected on the CG sites, and the forces predicted by the CG model, $\mathbf{F}^{CG} \in \mathbb{R}^{3N}$, for the N number of CG sites having coordinates $\mathbf{x}_i$:

$$\chi^2 = \langle \left[\mathbf{F}^{CG}(\mathbf{x}_i) - \mathcal{M}\left(\mathbf{F}^A(\mathbf{r}_j)\right)\right]^2 \rangle$$

$\mathcal{M}$ is the mapping operator between the atomistic and the CG space, $\mathbf{r}_j$ are the atomistic coordinates and the brackets $\langle \dots \rangle$ denote average over CG degrees of freedom and the number of sampled atomistic configurations.

This approach does not include structural information during the fitting, therefore, the ability of a CG model to represent the structure is a true test of its quality and an important characteristic that can be used for the comparison among different models.

## 2.2 Local environment representation and SchNet architecture

A key step in the development of ML methods for molecular modelling is the transformation of the coordinates of the system into input features for the ML model. In general, a transformation of the coordinates into a suitable descriptor must provide invariance with respect to translation, rotation, and permutation of the particles' order. Moreover, it is important to have a descriptor whose size does not depend on the number of particles in the system. Thereby, the resulting ML model is independent of system size and can be efficiently deployed to larger systems, while having been trained on data for systems of smaller size.

Rather than constructing ad hoc a suitable feature representation, more general and transferable strategies have been proposed, which consist of learning the local representation instead, together with the desired property, through a ML model. One pioneering effort in this regard is known as SchNet [6], and this idea was implemented utilizing a GCNN architecture. For a detailed description of the SchNet architecture, the following references are recommended: [6], [9]. The main features are recalled here briefly. The model was initially utilized for the migration from the quantum mechanical level to the atomistic one, but its application can be extended for the study of molecular systems at the CG level as well [10].

Each particle is represented through a feature vector, which is initialized to distinguish between the particle chemical identities (embedding layer). The feature is then updated in each neural network layer depending on the chemical environment, by performing continuous convolutions across the particle neighborhoods, optimizing the convolutional filter weights during the training (convolutional layers). Afterwards, a fully connected section is included (readout layers). A sequence of continuous convolution layers and readout layers constitutes a SchNet "block". Multiple blocks can be utilized in series to define the full network architecture. A schematic depiction of the procedure is given in Fig. 1.

The output found at the end of the blocks can be interpreted as a learned feature representation, which encodes the many-body information from the particle neighborhood required to predict the target property. Finally, a fully connected (dense) section transforms the fingerprints into the final scalar output, which is interpreted as a per-particle energy contribution. This local decomposition ensures invariance of the ML potential architecture to the total number of atoms. All energy contributions are summed to obtain the total energy ($\Sigma_i \widehat{U}_i$),, which is then differentiated with respect to the positions of the particles to predict the force acting on each particle. These are the target properties for the network training. During the network optimization the mean squared difference between the predicted forces and the forces associated with the molecular dynamics (MD) trajectory considered as input is minimized, using the following loss function $L$:

$$L = \frac{1}{3N}\sum_i^N \left\{\left[-\nabla_{\mathbf{x}_i}(\Sigma_i \widehat{U}_i + U_{ex})\right] - \mathcal{M}\left(\mathbf{F}^A(\mathbf{r}_j)\right)\right\}^2$$

where N is the total number of CG particles, $\mathcal{M}\left(\mathbf{F}^A(\mathbf{r}_j)\right)$ is the force acting on the CG particle at position $\mathbf{x}_i$, obtained from the atomistic MD simulation, mapped to the CG space, while the term in square brackets is the force acting on the particle at position $\mathbf{x}_i$ predicted by the GCNN. The term $U_{ex}$ included in the loss function is defined and explained in Section 2.3.

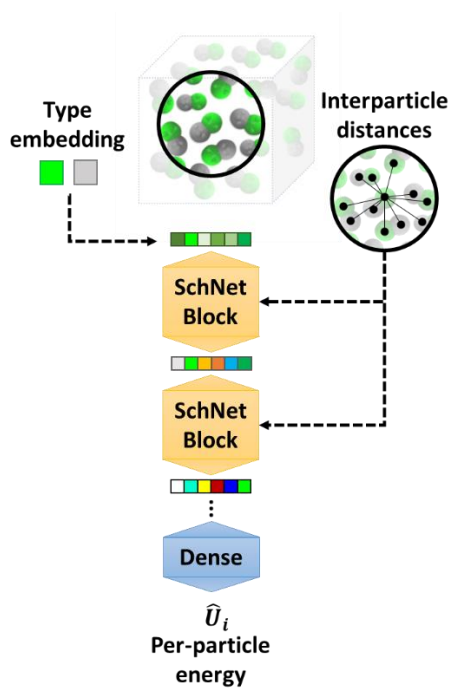

**Figure 1:** Schematic representation of the architecture for the per-particle energy prediction. The input information is the particle type and the interparticle distances in the neighborhood of each atom (within a cutoff radius).

## 2.3 Excluded volume prior

During the CG simulations with the trained ML force field, when parts of the coordinate space are reached that are very different from any point in the training set, the network is unable to reproduce the correct physical behavior which it has never been exposed to during training. In particular, the atomistic force field used to produce the training data ensures that the energy will become steeply very high when the configuration assumed by the system is departing from physical states, for example when atoms are moving too close to each other. These regions are not sampled in the underlying atomistic MD simulations, and therefore training data are missing for those regions of the configuration space. If such conditions occur during simulations performed with a neural network trained only on physically valid configurations, unphysical behavior can be produced, such as bonds overstretching, or atoms overlapping.

To mitigate this issue, several authors suggest to introduce regularization terms [10], related to non-bonded repulsion terms, to impose excluded volume effects, and intramolecular priors on specific geometric features, such as bonds and angles. The purpose of including such constraints is to ensure that the system energy will be driven to infinity if unphysical states occur that are not within the training data. For the system considered here, consisting of single-bead molecules, only excluded volume effects are relevant. In particular, an excluded volume energy term based on the pairwise distances between the moieties, is considered:

$$U_{ex} = \sum_{i=1}^{N-1} \sum_{j=i}^{N} \left( \frac{\sigma}{\|x_i - x_j\|} \right)^{n^{ex}}$$

where $\sigma$ and $n^{ex}$ are hyperparameters of the model, for which suitable values must be identified. This term is added to the total energy predicted by the GCNN, prior to differentiation.

## 3  RESULTS AND DISCUSSION

Tests were performed on three independent liquid benzene systems containing 500 molecules each, at 300 K (Fig. 2). The systems were initially equilibrated through a 1 ns NPT MD simulation, and then a 4 ns NVT run at the average equilibrium density was conducted. The first ns of the run was discarded, and, after that, 3000 configurations, saved every $10^4$ ps, were retained for each system, and constitute the training data for the force field development. Each benzene molecule was mapped into a single CG site. This choice allows to study the application of the ML method to a CG system that contains only intermolecular interactions, which are the most complex to represent and the ones for which the expressive power of a NN model could provide the greater advantage compared to traditional CG models.

NVT CG simulations with the ML potentials were conducted at 300 K using the ASE integrator [11].

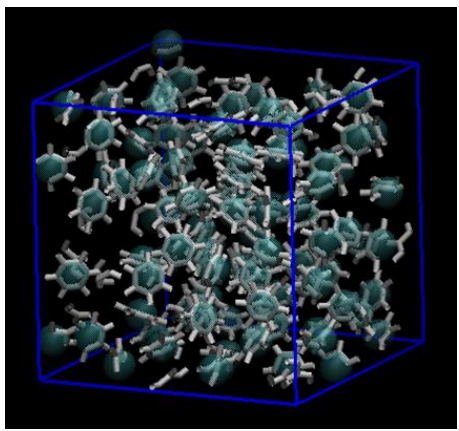

**Figure 2:** System studied: 500 molecules of liquid benzene at 300 K mapped onto one CG bead each (light blue spheres).

## 3.1 Hyperparameter values

The SchNet model contains several hyperparameters. In the following list, they are reported together with the values tested in this study:
- Feature size: 120, 240
- Convolutional filter size: 40, 80
- Number of filters: 128, 256
- Interaction blocks: 1, 2, 3
- Learning rate: $10^{-4}$
- Decay ratio: 0.5
- Batch size: 15
- Cut-off distance (Å): 7, 12
- Activation function: Tanh, Shifted Softplus
- Number of epochs: 100 – 2000
- Number of samples: 9000
- Excluded volume diameter $\sigma$ (Å): 3, 4, 5, 6, 7, 8, 9
- Excluded volume exponent $n^{ex}$: 5, 7, 9

80% of the data was used as a training set, while the remaining 20% as test set. The molecular configurations from the MD trajectory were randomly assigned to either set, and shuffled during the training.

Only few tests were conducted utilizing a cutoff of 12 Å. Compared to the case of 7 Å, models trained with this cutoff often exhibited overfitting behavior, with the loss over the test set becoming larger as the number of epochs increased. All results presented in the following sections were obtained with a cutoff of 7 Å. No hyperparameter optimization was performed for the learning rate, the decay ratio and the batch size. The effects manifested upon variation of the remaining hyperparameters are presented hereafter, through the comparison of selected results.

## 3.2 Model size and simulation stability

Several groups of tests were performed varying one of the hyperparameters related to the model size, namely either the feature size, filter size, number of filters, or number of blocks, to evaluate the effect on the training process and performance during the simulations. In the results shown in Fig. 3, the case in which the number of blocks is varied is shown. The following hyperparameters were constant across the tests presented in this section:
- Feature size: 240
- Convolutional filter size: 80
- Number of filters: 256
- Activation function: Shifted Softplus
- Excluded volume diameter $\sigma$ (Å): 5
- Excluded volume exponent $n^{ex}$: 7

The number of blocks was changed according to the legend in Fig. 3.

As can be observed in Fig. 3a, for tests A, B and C, the effect on the final loss of changing the number of blocks from 3 to 1 during training, is very limited. All tests reached a similar value of the loss function and this was observed also upon changing other hyperparameter values related to model size. This was observed also by Wang et al. [10], who reported the existence of a lower limit in the expected prediction error during the ML-based CG force field development procedure. The observed lower limit in each case is associated with the CG mapping level, and it is related to the multiplicity of atomistic configurations that map into the same CG configurations. Indeed, the CG level was the same for all the tests in this work.

When performing MD simulations with the trained models, very different behaviors are manifested (Fig. 3b). In particular, the larger models (higher number of blocks) exhibited lower kinetic energy fluctuations, while the smallest one had a divergent behavior: the very high temperature peaks correspond to fast in place oscillations of the molecules, which eventually fade, leaving the system in a "frozen" state. Husic et al. [12] suggested that the use of the hyperbolic tangent as activation function, instead of the shifted sotfplus one, could mitigate instabilities, however this was not the case for the system investigated in this study.

Case D is a test performed with the same hyperparameters as A, but different network initialization. In this case, the behavior was not reproducible between the two tests. Nonetheless, it seems that this is not a frequent occurrence: in the majority of cases the results are reproducible among tests performed with the same hyperparameters, and either they all produce stable behavior (non-divergence of the kinetic energy), or they all produce unstable behavior during the simulations (divergence of the kinetic energy), regardless of the network initialization.

It is interesting to include test D in the comparison because it shows the typical features of overfitting, with test set loss increasing as the training continues. Moreover, this test shows that there does not appear to be a clear correlation between the final loss level reached during the training and the amplitude of the kinetic energy oscillations.

These results imply that, in order to evaluate the fitness of a CG model, even roughly, it is not sufficient to solely observe the training output, but it is necessary to perform also a CG simulation with it, which is a computationally expensive step. On a positive note, it appears that the problematic behavior is manifested early on during the test CG simulations, therefore short runs suffice to retain or discard a trained model. Though not shown here, it was verified that the system that exhibited stability for these preliminary short simulations remained stable also during simulations 25 times longer. Moreover, independent simulations performed with the same NN force field were reproducible in terms of thermodynamic, structural and dynamical properties.

In addition to kinetic energy stability, it is possible to compare the simulated systems also in terms of structure, by computing the radial distribution functions between the centers of mass of the CG particles. The target distribution calculated from the atomistic data is shown in black in Fig. 2c. As can be seen, there is an inverse correlation between the amplitude of the kinetic energy oscillations and the emergence and sharpness of an unrealistic feature in the structure at approximately 3 Å. The corresponding molecular models show an uneven occupation of space, with void regions that are not expected in a liquid system. The origin of this feature or its relationship with the model architecture is not clear yet. Modifications to the method that improve the structure prediction will be presented in the section that follows. The same considerations made here about the effect of the number of blocks are valid also for the effect of the other hyperparameters related to model size.

## 3.3 Excluded volume prior effect

Fig. 4 reports the effect of the model hyperparameters related to the excluded volume prior, while keeping all the others constant:
1. Feature size: 240
2. Convolutional filter size: 80
3. Number of filters: 256
4. Activation function: hyperbolic tangent
5. Number of blocks 2

The excluded volume σ was changed according to the legend in Fig. 4, while three values of $n^{ex}$ are shown in the subfigures.

Changing the values of the excluded volume hyperparameters has a notable effect on the trend of the loss during training: higher $n^{ex}$ and higher σ require longer training to reach convergence in the loss value and, also, exhibit a less smooth trend of the loss during the optimization. Nevertheless, the final values of the loss still fall within a very narrow range. From all the models reported in Fig. 4, only two resulted in stable simulations, namely $n^{ex}$ = 7 and σ = 6 Å, and $n^{ex}$ = 9 and σ = 6 Å, hinting at the fact that possibly higher sigma values should be explored for this system. However, the systems simulated with these models still exhibited the unrealistic structural feature shown in Fig. 3c.

It was found that the structural representation can be improved if the loss function is modified to include also the squared deviation between the energy of the system predicted by the network and the intermolecular potential energy from the atomistic simulation, appropriately weighted with respect to the force component. As shown in Fig. 5, a model trained in this way with $n^{ex}$ = 7 and σ = 6 Å displayed stability in the kinetic energy and a more realistic structure. Moreover, consistent results were obtained by utilizing the model to simulate systems of different size than the one used for training, which is a necessary consistency trait for the exploitation of the CG potentials in simulating various system sizes, especially larger ones. Future efforts will be devoted to further investigating this promising direction and refining the results, also in comparison with predictions of previous methodologies for the development of CG potentials for the benzene systems that do not incorporate ML schemes [13]. To this aim, additional testing of the model hyperparameters that affect the structural representation will be necessary.

| Legend | A | B | C | D |
|---|---|---|---|---|
| Number of blocks | 3 | 2 | 1 | 3 |

| Legend | E | F | G | H | I |
|---|---|---|---|---|---|
| Excluded volume $\sigma$ (Å) | 2 | 3 | 4 | 5 | 6 |

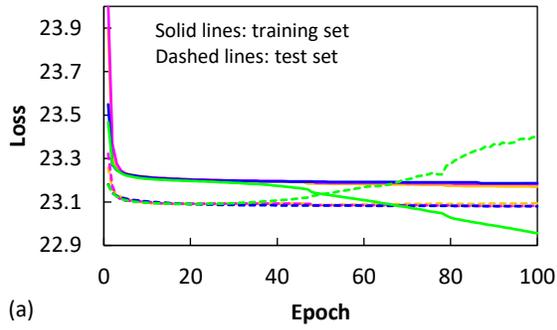

(a)

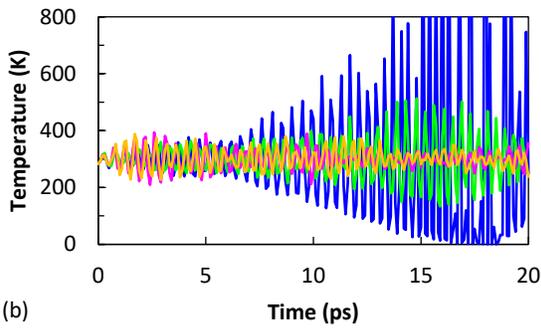

(b)

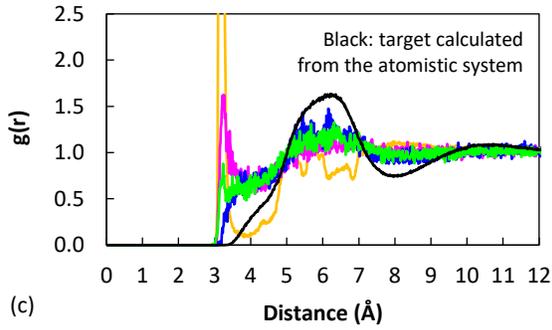

(c)

**Figure 3:** Effect of varying the number of blocks on (a) the final loss level, (b) simulation stability and (c) radial distribution function of the simulated system.

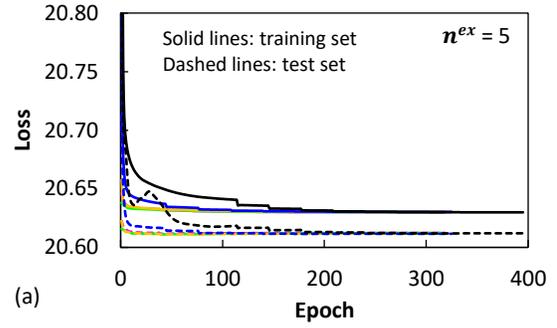

(a)

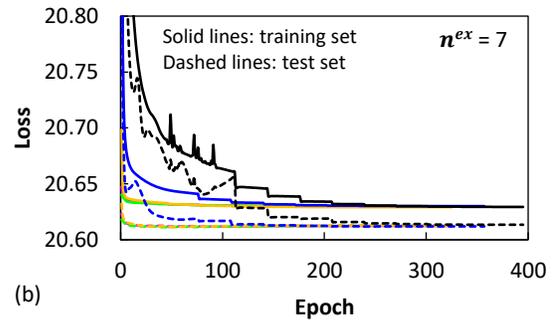

(b)

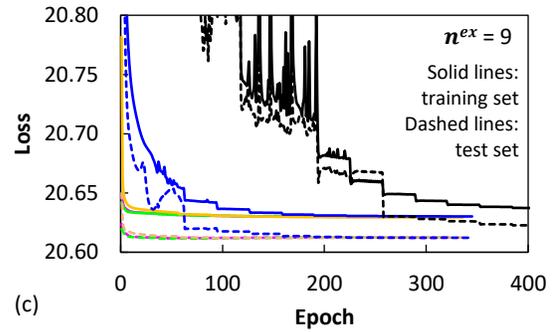

(c)

**Figure 4:** Effect of varying the excluded volume hyperparameters on the model training.

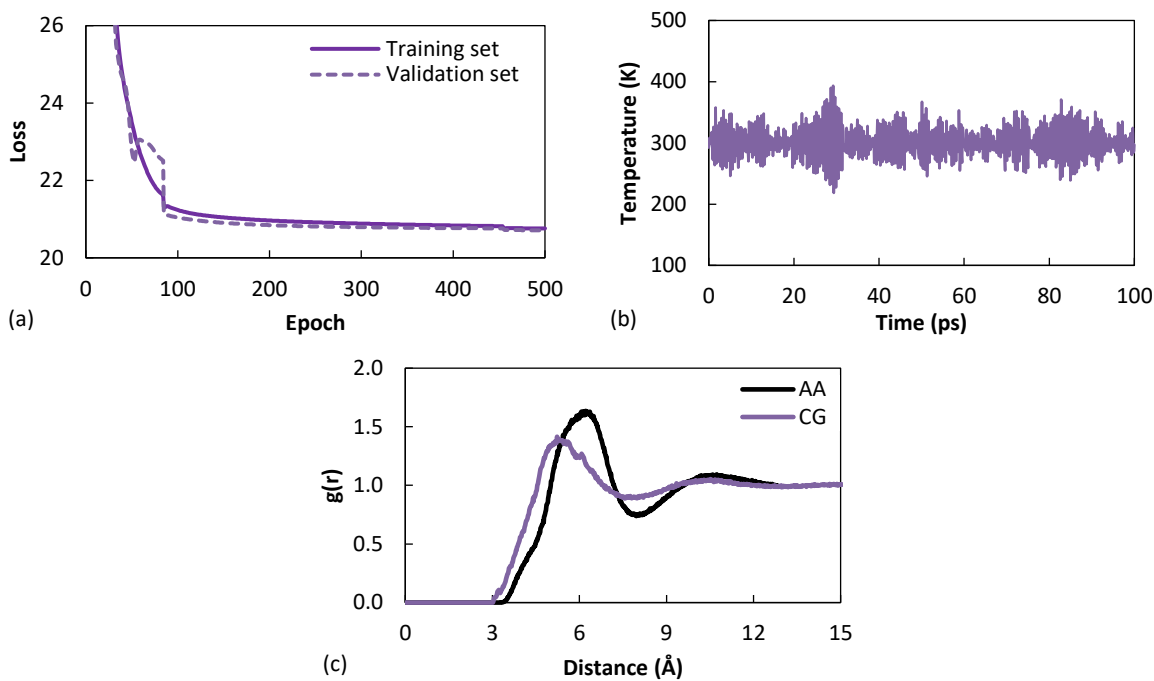

**Figure 5:** Results obtained with energy inclusion in the loss function for (a) training performance, (b) stability of the kinetic energy during the simulation and (c) structure.

## 4 CONCLUSIONS

In this work, the application of a powerful ML architecture was implemented for the generation of CG potentials of a bulk liquid system, *i.e.* benzene in a single-bead representation. We conducted a systematic investigation of the effect of the model hyperparameters on the results and highlighted challenges related to the difficulty of assessing the efficiency of a trained model solely from the performance during training. In fact, models trained with different hyperparameters can reach a very similar value for the loss function, and yet exhibit very different and in many cases unphysical behavior in terms of structure and/or dynamics during CG MD simulations.

Extensive search of the hyperparameters space is necessary to identify a suitable combination, which may be system specific. It would be highly beneficial, in terms of computational effort required to generate a viable model, if part of this search could be avoided. In this regard, it is envisioned that the introduction of the excluded volume prior could be bypassed by enriching the training dataset with samples of unphysical configurations. The appropriate generation and incorporation of such instances in the model might prove to be a not trivial task and cause new challenges to emerge.

This specific area of research and application is nascent and few reports exist about the application of ML CG models in molecular simulations, therefore it is of great importance to reveal and discuss, along with the advancements, also the difficulties, challenges and unexpected or problematic behavior that may be observed. Disclosing such information will aid the interdisciplinary scientific community to accumulate valuable knowledge and experience, in order to engage in the search for improved strategies that will accelerate discovery and expand successful applications.

**ACKNOWLEDGMENTS**

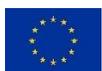
E.R. acknowledges funding from the European Union's Horizon 2020 research and innovation programme under the Marie Skłodowska-Curie grant agreement No 101030668. This work was supported by computational time granted from the National Infrastructures for Research and Technology S.A. (GRNET S.A.) in the National HPC facility - ARIS - under the project MULTIPOLS (ID: 011032).